\documentclass[sigconf]{custom-arxiv}  % do not change this line!

\usepackage{booktabs}

\usepackage{flushend}

\usepackage{amsmath,amssymb,amsfonts}
\usepackage{amsthm}
\usepackage{algorithm2e}
\usepackage{graphicx}
\usepackage{textcomp}
\usepackage{xcolor}
\usepackage{rotating}
\usepackage{tabularx}
\usepackage{pgfplots}
\pgfplotsset{compat=1.8}
\usepgfplotslibrary{statistics}

\begin{document}

\title{Culture-Based Explainable Human-Agent Deconfliction} 

\author{Alex Raymond, Hatice Gunes, and Amanda Prorok}
\affiliation{%
 \institution{Department of Computer Science and Technology \\ University of Cambridge \\ \{ar968, hg410, asp45\}@cam.ac.uk}
}

\begin{abstract}  
Law codes and regulations help organise societies for centuries, and as AI systems gain more autonomy, we question how human-agent systems can operate as peers under the same norms, especially when resources are contended. We posit that agents must be accountable and explainable by referring to which rules justify their decisions. The need for explanations is associated with user acceptance and trust. This paper's contribution is twofold: \textit{i)} we propose an argumentation-based human-agent architecture to map human regulations into a \textit{culture} for artificial agents with explainable behaviour. Our architecture leans on the notion of argumentative dialogues and generates explanations from the history of such dialogues; and \textit{ii)} we validate our architecture with a user study in the context of human-agent path deconfliction. Our results show that explanations provide a significantly higher improvement in human performance when systems are more complex. Consequently, we argue that the criteria defining the need of explanations should also consider the complexity of a system. Qualitative findings show that when rules are more complex, explanations significantly reduce the perception of challenge for humans.
\end{abstract}


\maketitle

%%%%%%%%%%%%%%%%%%%%%%%%%%%%%%%%%%%%%%%%%%%%%%%%%%%%%%%%%%%%%%%%%%%%%%%%%%%%%%%%%%%%%%%%%%%%%%%%%%%%%%%%%
%% start of main body of paper

\newcommand{\acronym}{X-CORE}

\section{Introduction}

The Code of Ur-Nammu is the oldest known written law code, inscribed around 2100 BC in ancient Mesopotamia \cite{Finkelstein1968TheUr-Nammu}. Its structure is a set of rules (carved in a stone tablet) designed to aid denizens to settle potential conflicts. Conceptually, little has changed since then, as humans historically and currently rely on sets of rules to specify their own systems' behaviours, expecting peers to abide by those regulations when conflicts arise. Different regimens are defined for several environments, be it traffic, competitive sports, business, civil society, etc.

Inasmuch as robots and intelligent artificial agents progress in sophistication, they obtain increasingly more autonomy and start taking part in the same systems and societies that humans do, no longer as tools, but rather as peers. It is therefore fundamental to guarantee that agents embedded in those environments will also observe and respect the same rules and regulations that humans do in order to resolve conflicts and operate orderly \cite{Benjamin2006NavigationRoad}. The need for rule-abiding behaviour goes beyond the debate of ethics and morals \cite{Pagallo2016EvenLaw}, which is vast but out of the scope of this investigation. Instead, we are interested in ensuring that autonomous agents consider their \textit{liability} \cite{Asaro2016TheAgents} and can express justification for their agency \cite{Langley2017ExplainableSystems} \textit{with regards to the present ruleset to be followed} --- preferably in a human-understandable way.

\citet{Rizaldi2015FormalisingVehicles} tackle the liability and accountability problem in the autonomous vehicle domain with a manual formalisation of specific traffic rules, using automated theorem proving techniques. Their approach is formally rigorous but is specialised for a specific set of traffic rules only and does not generalise beyond. \citet{Cranefield2019AccountabilityAgents} propose that ideal accountable agents must: \textit{i)} understand what is expected from them (from rules/obligations); \textit{ii)} answer queries about their decision-making (being explainable); \textit{iii)} carry out argumentative dialogues in which beliefs and plans are challenged and justified; \textit{iv)} adapt their reasoning apparatuses or update their plans as a result of accountability dialogues; and \textit{v)} take human values into account when reasoning.

The quest for \textit{understanding} and \textit{explaining} the decisions made by artificially intelligent systems and agents motivated the materialisation of the \textit{eXplainable AI} (XAI) \cite{Gunning2017Explainablexai} research field. The onset of machine learning systems and the popularity of methods such as support vector machines and artificial neural networks have led to AI solutions that are efficient but indecipherable regarding their rationale behind a conclusion. For that reason, the XAI community is interested in systems that are not only clear regarding \textit{`how'}, but also as to \textit{`why'} certain decisions were made \cite{Rosenfeld2019ExplainabilitySystems, Anjomshoae2019ExplainableReview}.

In order to achieve more realistic explainability for humans, we spur the necessity for more realistic models of reasoning. Expressly, classical logic does not provide an authentic representation of common sense reasoning, as under a scenario of incomplete information a human may draw conclusions that can be withdrawn later, when new information is presented \cite{Amgoud2009UsingDecisions}. Argumentation-based approaches attempt to fill in this gap by providing a framework for defeasible reasoning \cite{Dung1995OnGames}, which grants systems clear decision-making mechanisms that provide not only resolutions, but also the \textit{reasons} that may support it \cite{Zeng2018BuildingArgumentation}. 

Argumentation approaches walk hand in hand with the desiderata proposed by \citet{Cranefield2019AccountabilityAgents}, as they allow us to: \textit{i)} enable norm-aware reasoning  \cite{Bench-Capon2019NormsFrameworks}; \textit{ii)} generate explanations \cite{Sklar2018ExplanationArgumentation, Fan2015OnArgumentation}; \textit{iii)} carry argumentative dialogues to support their positions \cite{Rosenfeld2016StrategicalPersuasion, Amgoud2009UsingDecisions}; \textit{iv)} perform meta-reasoning \cite{Young2018InstantiatingFrameworks}; and \textit{v)} consider human values \cite{Kakas2003ArgumentationAgents}. Consequently, we regard argumentation frameworks as a strong mechanism for providing accountable and explainable agency.

\citet{Rosenfeld2019ExplainabilitySystems} postulate that the \textit{necessity} for explainability in human-agent systems follows a taxonomy of three types of explanations: \textit{not helpful}, \textit{beneficial}, and \textit{critical}. They posit that if humans will not accept a system without an explanation, then the need for explainability is critical. Likewise, explanations can range in significance depending on their ability to engender trust in human users. We aim to introduce another dimension to this analysis, by empirically observing that the complexity of the rules governing a system may also affect the usefulness of explanations where human performance is concerned.

In this work, we address environments where humans and agents act with independent agency and are subjected to the same rules and conditions. Most explainable approaches in human-agent systems are classified with regards to their  human-centric or agent-centric \cite{Rosenfeld2019ExplainabilitySystems} approaches, but relatively few are interested in emulating human-agent societies \cite{Billhardt2014AnSocieties}. Can agents and humans with individual goals coexist as peers in a norm-aware environment where resources are limited? Can such peers resolve conflicts and provide accountability to their decisions, to both humans and other agents alike? Namely, given a multi-agent environment with resource contention, can we define a mechanism that allows us to facilitate human-agent integration by providing: \textit{i)} an equivalence between human-readable rulesets and agent policies and \textit{ii)} in a way that is explainable and allows humans to interact successfully with agents to resolve conflicts?

This paper offers the following contributions:
\begin{itemize}
    \item We propose an argumentation-based architecture for designing explainable human-agent systems for deconfliction environments.
    \item We present an empirical study to investigate the effect of explanations in this architecture in varying levels of complexity.
\end{itemize}

We exemplify our architecture with a multi-agent resource contention application in the context of the problem of \textit{multi-agent path deconfliction}. We show how humans and agents can deconflict trajectories whilst respecting externally-defined ``rules of way.''

Towards this end, we design a computer game implementing the proposed architecture and conduct a user study to evaluate it. Humans are given path deconfliction rulesets with different amounts of rules each and are asked to navigate in a multi-agent environment and avoid collisions with agents. In our setting, we define complexity as the number of rules that govern the deconfliction of resources. We observe how humans perform in terms of ruleset complexity and the presence/absence of explanations. Our results show that the benefit of explanations is correlated with the complexity of the underlying system. Qualitative results show that human experience in systems with explanations is superior when such systems are sufficiently complex.

\section{Background}

In this section, we introduce definitions and concepts that are used in the construction of our architecture. Section~\ref{section:abstract-argumentation} introduces essential definitions for Abstract Argumentation frameworks, the principal \textit{deliberation} tool in our framework. Section~\ref{section:dialectic} presents a mechanism for dialogical exchanges between agents in Abstract Argumentation, followed by an argumentation-based formalism of explanations in Section~\ref{section:explanations}.

\subsection{Abstract Argumentation}
\label{section:abstract-argumentation}

A seminal paper from Dung \cite{Dung1995OnGames} introduces the concept of an \textit{argumentation framework}, also called \textit{abstract argumentation} (AA). His framework considers arguments as purely abstract entities, with no special attention paid to their internal structure. Modelling occurs at the level of relationships between those abstract entities.

The main concept behind AA is that a statement is acceptable if it can be defended successfully against attacking arguments. As put by Bentahar et al.~\cite{Bentahar2010ARepresentation}, \textit{`the beliefs of a rational agent are characterised by the relations between its ``internal arguments'' supporting its beliefs and the ``external arguments'' supporting contrary beliefs.'}

We will use and adapt some definitions from Dung's work and other authors~\cite{Coste-Marquis2005SymmetricFrameworks, Modgil2014RevisitingFrameworks,Modgil2009ProofFrameworks}, as follows.

\begin{definition}
An \emph{argumentation framework} is a directed graph $AF = (\mathcal{A}, \mathcal{R}),$ where $\mathcal{A}$ is a set of arguments (vertices) and R is a set of directed, binary attack relationships between arguments (arcs), i.e., $\mathcal{R} \subseteq \mathcal{A} \times \mathcal{A}$. We also say $\textit{attacks}(a,b)$ holds iff $(a,b) \in \mathcal{R}$. Likewise, a set $S$ of arguments attacks another set of arguments $T$ (or $T$ is attacked by $S$) if any argument in $S$ attacks an argument in $T$.
\end{definition}

\begin{definition}
An argument $a \in \mathcal{A}$ is \textit{acceptable} with respect to a set $S$ of arguments iff for each argument $b \in \mathcal{A}$ that attacks $a$ there is a $c \in S$ that attacks $b$. In that case, $c$ is said to \textit{defend} $a$.
\end{definition}

\begin{definition}
A set of arguments $S$ is said to be \textit{conflict-free} if there is no attack within its arguments, i.e. there are no arguments $a, b \in S$ s.t. $a$ attacks $b$. Likewise, a $S \subseteq \mathcal{A}$ of arguments is \textit{admissible} iff it is conflict-free and each argument in $S$ is acceptable with respect to $S$.
\end{definition}

\subsection{Dialogue Game Rules}
\label{section:dialectic}

The extension semantics introduced by Dung are powerful in asserting global properties of the argumentation framework, but their output is static and monological in nature. 

In pursuance of a more dialogical approach \cite{Karamlou2019DecidingArgumentation}, one must consider the dynamics of dialogue and the assumptions therewithin. Using Jakobovits and Vermeir's position framework formalism~\cite{Jakobovits1999DialecticFrameworks}, \textit{`the combination of a set of rules that govern the game, and the determination of winning criteria, constitute a dialectic semantics for the ``theory'' that underlies the player's arguments.'} We will adapt some of the definitions from~\cite{Jakobovits1999DialecticFrameworks}, as follows.

\begin{definition}
A \textit{position framework} paired with an argumentation framework $AF = (\mathcal{A}, \mathcal{R})$ is a position framework $PF = (\mathcal{P}, \mathcal{R}^*)$, where $\mathcal{P}$ consists of conflict-free subsets of $\mathcal{A}$, and $\mathcal{R}^*$ denotes the set of finite sequences of elements from $R$. Elements of $\mathcal{P}$ are called \textit{positions}.
\end{definition}

\begin{definition}
A \textit{player} $c$ can be categorised as the \textit{proponent} ($p$) or \textit{opponent} ($o$). The adversary of $p$ is denoted $\overline{p} = o$. Conversely, $\overline{o} = p$. 
\end{definition}

\begin{definition}
Let a player $c \in \{p, o\}$ and a position $X \in \mathcal{P}$. A \textit{move} in $\mathcal{P}$ is a pair $(c, X)$. For a move $m = (c, X)$, we use $\textit{player}(m)$ to denote $c$ and $\textit{pos}(m)$ to denote $X$.
\end{definition}

\begin{definition}
\label{definition:dialogue-type}
A \textit{dialogue type} is a tuple $(\mathcal{P}, \mathcal{R}^*, \phi)$, where $(\mathcal{P}, \mathcal{R}^*)$ is a position framework and $\phi : \mathcal{P}^* \xrightarrow{} 2^\mathcal{P}$ is a \textit{legal-move} function. A \textit{dialogue} $D$ in $(\mathcal{P}, \mathcal{R}^*, \phi)$ is any countable sequence $d_0,d_1,\dots,d_n$ of moves in $\mathcal{P}$ that satisfies:
\begin{enumerate}
    \item $\textit{player}(d_{i+1}) = \overline{\textit{player}}(d_i)$, i.e. the players take turns.
    \item $\textit{pos}(d_{i+1}) \in \phi(\textit{pos}(d_0)\dots\textit{pos}(d_i))$, i.e. the next move is legal.
    \item $d_{i+1} \notin \{d_0, d_1, \dots, d_i\}$, i.e. a move cannot be repeated twice.
    \item $\textit{attacks}(\textit{pos}(d_{i+1}), \textit{pos}(d_i))$, it attacks the adversary's last move
    \item $\textit{player}(d_0) = p$, i.e. the proponent makes the first move.
\end{enumerate}

The dialogue $D$ is said to be \textit{about the position} $\textit{pos}(d_0)$.
\end{definition}

\begin{definition}
Let $X^\gets$ and $X^\to$ denote the sets of positions that attack and are attacked by $X \in \mathcal{P}$, respectively. A player $c$ is said to \textit{win} $D$ if $D$ is finite and ends with a move $(c, X)$ s.t. $X^\gets{} \cap \phi(D) = \emptyset$, i.e., the dialogue cannot be continued.  
\end{definition}

\begin{definition}
Let $(\mathcal{P}, \mathcal{R}^*)$ be a position framework. The legal-move function $\psi_{(\mathcal{P}, \mathcal{R}^*)} : \mathcal{P}^* \xrightarrow{} 2^\mathcal{P}$ which allows non-self-defeating nor useless moves in $(\mathcal{P}, \mathcal{R}^*)$ is defined as follows: $\forall Y_0,\dots, Y_i$~$\in$~$\mathcal{P}^*$:

$$\phi_{(\mathcal{P}, \mathcal{R}^*)}(Y_0, \dots, Y_i) = \mathcal{P}\setminus(\{X \mid \overbrace{\textit{attacks}(X, X)}^\text{self-defeating}\} \cup \overbrace{\bigcup\limits_{j=o}^{i} Y_j^\to}^\text{useless})$$
\end{definition}

We now have sufficient tools to formalise types of dialogues that encompass the previously chosen rules, with single or multiple arguments per move:

\begin{definition}
Let $AF = (\mathcal{A}, \mathcal{R})$ be an argumentation framework. A \textit{useful-single-argument dialogue} in $AF$ is a dialogue in the dialogue type $(\mathcal{A}', \mathcal{R}, \psi_{(\mathcal{A}', \mathcal{R})})$, where $\mathcal{A}' = \{\{a\} \mid a \in \mathcal{A}\}$ and $\psi_{(\mathcal{A}', \mathcal{R})}$ designates moves that are not self-defeating nor useless. The dialogue type $(\mathcal{P}, \mathcal{R}, \psi_{(\mathcal{P}, \mathcal{R})})$ is called the \textit{useful-multiple-argument} dialogue type, where $\mathcal{P}$ is the set of conflict-free subsets of $\mathcal{A}$.
\end{definition}

\subsection{Explanations}
\label{section:explanations}

\citet{Fan2015OnArgumentation} propose an argumentation semantics aimed at generating explanations. This formalism promotes the notion of explanations as sets of arguments, taking into consideration which arguments contribute to the justification (or \textit{r-defence}) of a specific premise (argument). We utilise their definitions for our framework, as follows.

\begin{definition}
Given an AA framework $AF = (\mathcal{A}, \mathcal{R})$, let $a, b \in \mathcal{A}$. $a$ \textit{r-defends} $b$ iff:
\begin{enumerate}
    \item $a = b$; or
    \item $\exists z \in \mathcal{A}$, s.t. $a$ attacks $z$ and $z$ attacks $b$; or
    \item $\exists z \in \mathcal{A}$, s.t.~ $a$ r-defends $z$ and $z$ r-defends $b$.
\end{enumerate}

$S \subseteq \mathcal{A}$ \textit{r-defends} $a \in \mathcal{A}$ iff $\forall b \in S$: $b$ r-defends $a$.
\end{definition}

\begin{definition}
A set of arguments $S \subseteq \mathcal{A}$ is \textit{related admissible} iff $\exists a \in S$ s.t. $S$ r-defends $a$ and $S$ is admissible. $a$ is said to be a \textit{topic} of $S$. For any argument $a \in \mathcal{A}$, an \textit{explanation} of $a$ is $S \subseteq \mathcal{A}$ s.t. $S$ is a related admissible set and $a$ is a topic of $S$.
\end{definition}

Their definition of explanations is further characterised by a classification with regards to cardinality and set inclusion:

\begin{definition}
Let $a \in \mathcal{A}$ and $E_a$ be the set of all possible explanations of $a$. For every $S \in E_a$, we say $S$ is a \textit{minimal} or \textit{maximal} explanation iff $S$ is the smallest or largest subset of $E_a$ with regards to cardinality, respectively. Similarly, $S$ is a \textit{compact} or a \textit{verbose} explanation iff $S$ is the smallest or largest subset of $E_a$ with regards to set inclusion, respectively.
\label{definition:toni-explanation}
\end{definition}

\section{Proposed Architecture}

We introduce an architecture for \textit{explainable conflict resolution} (\acronym) as a mechanism that provides explainable deliberation capabilities for dialectic interactions between agents. Below, we elaborate on definitions and concepts that were created for the purpose of this application.

\begin{example}
Suppose the following situation: vehicle $A$ crosses a green light in a junction and is about to collide with vehicle $B$, who ran a red light. In most highway codes, the rule \textit{`a vehicle shall not cross the stop line on a red light'} can be ignored if rule \textit{`a vehicle may cross the stop line on a red light if it is an emergency vehicle'} also applies to that situation. Therefore, in this specific situation, the right of way can be determined by $B$'s status: if it is an emergency vehicle, then it could refer to the aforementioned rule and \textit{argue} in favour of its right of way. Likewise, if $B$ is not an emergency vehicle, then it would not be able to defeat $A$'s claim of the first rule and $B$ would find itself at fault.
\end{example}

Consequently, what would happen if either $A$ or $B$ is an artificial autonomous agent? Would the human counterpart benefit more from being explicitly told which rules are being used, or is having an implicit knowledge sufficient? We can rather evidently demonstrate that there is no challenge in having autonomous agents follow rule-based systems. Instead, our architecture aims to create a direct mapping between rulesets in human-readable form and corresponding argumentation frameworks.

For our problem domain, we assume a setting where agents perform localised decision-making. When acting, we are interested in ensuring that each agent's behaviour is compliant with an overall \textit{culture} (represented by an argumentation framework $AF$) shared amongst all participants in the system. In order to check which rules apply in a specific event of a conflict, we introduce a mechanism of \textit{argument verification} (Section~\ref{section:verification}). Finally, after agents and humans share a common model and can provide evidence for their rule-compliant justification, we demonstrate how to build explanations from this framework in Section~\ref{section:explanation-generation}.
 
\subsection{Culture}

Orderly behaviour can only happen if all agents share common guidelines and understand the same rules. We define the notion of \textit{culture} as a collective agreement of norms and priorities, represented by an argumentation framework.

\begin{definition}
Let any two players $p, o$ be the proponent and opponent in a dialogue game. We say a \textit{proposition} is any argument $a \in \mathcal{A}$ that may be used by proponent $p$ to request a contended resource from opponent $o$.
\end{definition}

\begin{definition}
Let $\mathcal{K} \subseteq \mathcal{A}$ be the set of all propositions in $\mathcal{A}$. We say a system has a \textit{culture} $C = (\mathcal{A}, \mathcal{R}, \mathcal{K})$ iff $|\mathcal{K}| > 0 $ and all agents share $C$ as their culture.
\end{definition}

\begin{example}
\label{example:ambulance}
A simple example would be: suppose a culture that contains three arguments, $\mathcal{A} = \{\mu, \alpha, \beta\}$, where $\mu$ represents the proposition \textit{`I have right of way'}, $\alpha$ represents \textit{`I am an ambulance`} and $\beta$ represents \textit{`I am a fire rescue truck`}. Defining $\mathcal{R}_a = \{(\alpha, \mu), (\beta, \mu), (\alpha, \beta)\}$ is akin to defining that, in this application, ambulances have priority over fire trucks. Conversely, defining $\mathcal{R}_b = \{(\alpha, \mu), (\beta, \mu), (\beta, \alpha)\}$ would mean the opposite. Despite having the same argument set $\mathcal{A}$, $C_a = (\mathcal{A}, \mathcal{R}_a, \mathcal{K})$ is a different culture from $C_b = (\mathcal{A}, \mathcal{R}_b, \mathcal{K})$. 
\end{example}

\subsection{Propositional Dialogues}

When agents are presented with conflicts that require a compliant resolution (with regards to the ruleset), a dialogue game starts from the proponent $p$. Each agent then takes turns in choosing arguments that are potentially able to defeat the previous, as shown in Definition~\ref{definition:dialogue-type}. Agents can use one or multiple arguments at each turn depending on whether it is a \textit{useful-single} or \textit{useful-multiple-argument} dialogue type. The game ends when one agent provides an argument that cannot be defeated by any of the other agent's possible arguments and thus has to concede or reject the initial proposition, depending on the result. 

We extend the set of requisites for a dialogue seen in Definition~\ref{definition:dialogue-type} and propose the idea of a \textit{propositional dialogue}:
\begin{definition}
Let $D = \{d_0, \dots, d_i\}$ be a dialogue in a position framework $PF = (\mathcal{P}, \mathcal{R}^*, \phi)$ paired with a culture $C = (\mathcal{A}, \mathcal{R}, \mathcal{K})$. We say $D$ is a \textit{propositional dialogue} iff $D$ is about a position $pos(d_0)$ where $\exists a \in pos(d_0)$ s.t. $a$ is a proposition.
\end{definition}

\begin{definition}
Let $D = \{d_0, \dots, d_i\}$ be a propositional dialogue. We denote $d_0$ as the \textit{motion} of the dialogue.
\end{definition}

The player who wins $D$ then takes priority or ownership with regards to the contended resource disputed in the proposition. However, as cultures may be constant, they need to cater to most circumstances in the environment. In Example~\ref{example:ambulance}, the ability of an agent-player $c$ using $\alpha$ as an argument to defeat $\mu$ depends exclusively on the fact of $c$ being, in fact, an ambulance. We introduce the concept of \textit{argument verification} to deal with this matter.

\subsection{Argument Verification}
\label{section:verification}

Most applications of argumentation frameworks consider frameworks as static, i.e., the combination of all the arguments may generate an \textit{extension} or \textit{labelling} that represents an insight about which arguments should or should not be admitted. In our case, the culture denotes a ruleset that does not account for a specific happenstance, but rather a more comprehensive model that encompasses different future scenarios. For that purpose, every agent has to \textit{verify} which arguments are valid at a particular moment. 
We propose the architecture of \textit{argument verification} to address the issue of factual correctness and \textit{validity} of a specific argument given the context of its proponent. 

The verification of arguments can be modelled as decision problems:

\begin{definition}
Let $\alpha \in \mathcal{A}$ and $c \in \{p,o\}$ be an argument and a player, respectively. We denote $\alpha$ as \textit{demonstrable} by agent-player $c$ iff checking the correctness of that argument admits a finite and computable decision procedure.
\end{definition}

We can naturally extend this definition to encompass the complexity of argument verification. For example, \textsf{P}\textit{-demonstrable arguments} represent arguments whose associated decision problem is in \textsf{P}.

This decision procedure can be represented by a predicate function that evaluates, in the current context, whether a specific argument may be used or not.

\begin{definition}
Let $\zeta$ denote the set of all possible contexts in the environment. $\forall \alpha \in \mathcal{A}$, $\alpha$ admits a predicate function $f_\alpha: c, z \rightarrow \{\texttt{True}, \texttt{False}\}$, where $c$ represents the player and $z \in \zeta$ is a context. We say $f_\alpha$ is the \textit{verifier} function of argument $\alpha$. A special case applies for propositions, as they are hypothetical and cannot be checked for correctness.
\end{definition}

\begin{corollary}
The verifier function of every proposition always returns $\texttt{True}$.
\end{corollary}

\begin{definition}
Let $\alpha \in \mathcal{A}$ be an argument. Let $c \in \{p,o\}$ be a player and $z \in \zeta$ a context. We say argument $\alpha$ is \textit{demonstrably true} by player $c$ iff $f_\alpha(c, z) = \texttt{True}$.
\end{definition}

\begin{definition}
Let $D = \{d_0, \dots, d_i\}$ be a dialogue. Let $c \in \{p,o\}$ be a player and $X \in \mathcal{P}$ be a position. We say $d_{i+1} = (c, X)$ is a \textit{verified move} iff $\forall a \in pos(d_{i+1})$, $f_a(c) = \texttt{True}$. 
\end{definition}

Note that demonstrably true arguments do not mean that they are universally true -- not even that they are true at all. All it means is that an agent will be able to compute a procedure to check if that statement stands against its own knowledge in the current context. The notion of demonstrably true is, in fact, a local definition of truth, as it only requires the perception of a single agent, even if the agent is mistaken/uneducated about the world (such as in systems with imperfect/incomplete information.)

\begin{figure}
\includegraphics[width=8.7cm]{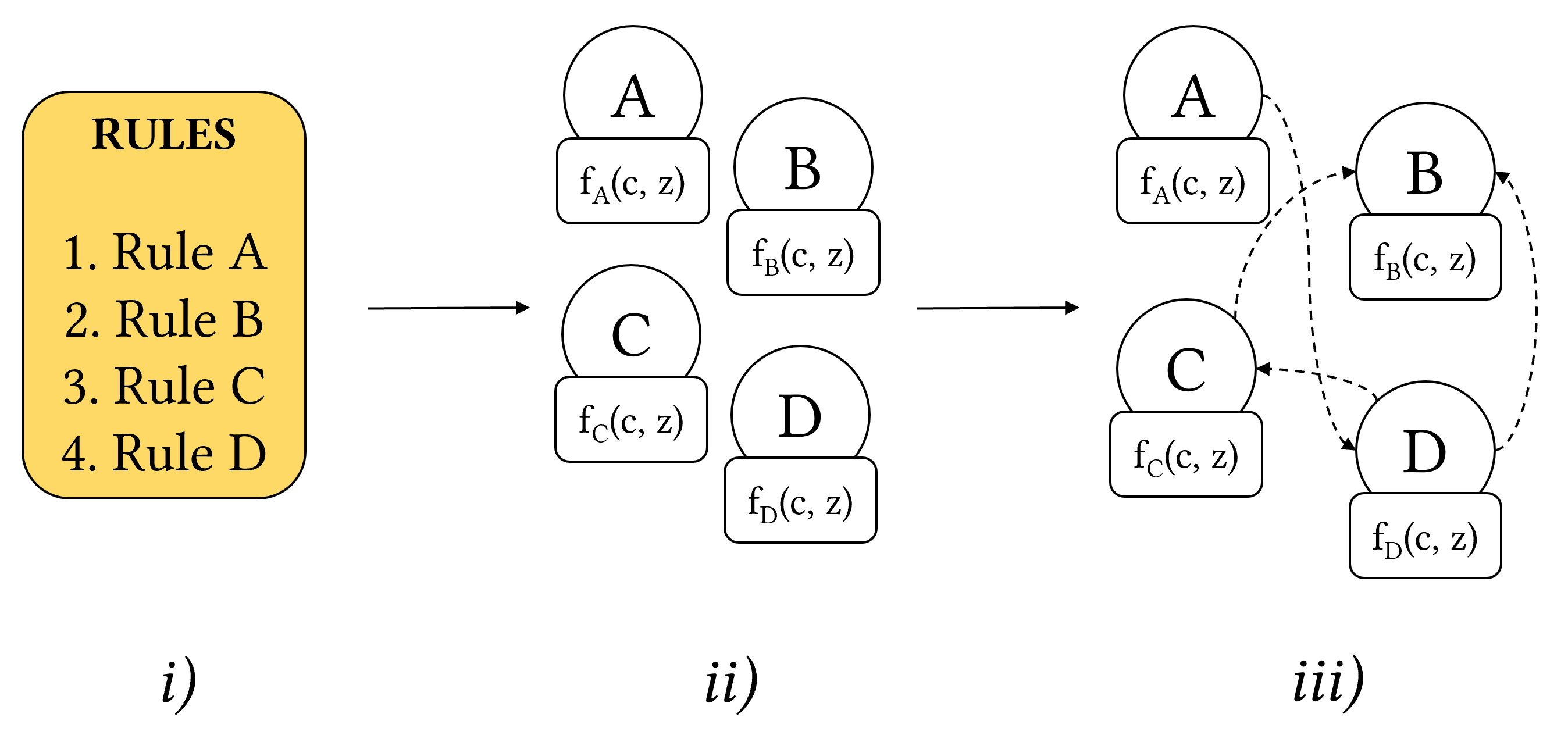}
\caption{Example diagram of \acronym: \textit{i)} Ruleset; \textit{ii)} Mapping rules into pairs of arguments and verifier functions; \textit{iii)} Defining attacks between arguments.}
\label{fig:rulestoarguments}
\end{figure}

All deliberation in this system is delegated into the specifics of the predicate functions that accompany each argument in the system. Hence, the design of a system in \acronym~ is divided in two phases (see Figure~\ref{fig:rulestoarguments}): 

\begin{itemize}
    \item Mapping rules into pairs of arguments and verifier predicate functions;
    \item Establishing attack relationships between generated arguments.
\end{itemize}

As every rule is represented by an argument, tracing the history of exchanged arguments in this manner provides an insight on each agent's attempt to justify their prioritisation based on the ruleset provided. We facilitate the interaction with humans by providing \textit{explanations} to justify the results of the dialogue game.

\subsection{Explanation Generation}
\label{section:explanation-generation}

\acronym~ does not bind agents into a specific strategy for choosing moves in a dialogue game. The justification is that humans cannot be bound to a unique way of thinking, or be \textit{programmable} as an artificial agent can. Therefore, we allow agents to freely choose their (verified) moves and focus on generating \textit{post-hoc} explanations derived from the history of a dialogue $D$. For that purpose, we propose some mechanisms for generating explanations in \acronym, based on the definitions seen in \cite{Fan2015OnArgumentation}.

\begin{definition}
Let $D = \{d_0, \dots, d_i\}$ be a completed dialogue and $d_i$ be the \textit{winner} move. We say a set $W \subseteq D$ is the set of \textit{winning} moves where $W = \{ d_k \in D$ $|$ $ player(d_k) = player(d_i) \}$. The set of \textit{losing} moves is denoted by $L = D \setminus W$.
\end{definition}

\begin{definition}
Let $d_i$ be the winner move in $D$. An \textit{explanation} $E_D$ of $D$ is defined as $E_D \subseteq D$ s.t. $\exists X = d_i$ and $X \in E_D$, i.e., it always contains the winner move. We denote $\mathcal{E}_D$ as the set of all possible explanations of a dialogue $D$.
\end{definition}

We can create a notion of \textit{contrastive explanations} to include losing moves. The idea behind contrastive explanations is to provide extra justification as to why a specific argument was \textit{not} accepted. We denote explanations without losing moves as \textit{plain explanations}.

\begin{definition}
Let $d_i$ be the winning move in $D$. A \textit{contrastive} explanation $CE_D$ of $D$ is defined as $CE_D \in \mathcal{E}_D$ s.t. $\exists X = d_i, \exists Y \in L$, and $X,Y \in CE_D$. A \textit{plain} explanation $PE_D$ of $D$ is defined as $PE_D \in \mathcal{E}_D$ s.t. $\forall X \in PE_D$, $X \in W$.
\end{definition}

\begin{definition}
Adapted from Definition \ref{definition:toni-explanation}. Let $\mathcal{E}_D$ be the set of all possible explanations of $D$. We therefore say that, for any $S \in \mathcal{E}_D$, $S$ is a: \textit{minimal} or \textit{maximal} explanation iff $S$ is a smallest or largest subset of $\mathcal{E}_D$ with regards to cardinality, respectively. $S$ is a \textit{compact} or a \textit{verbose} explanation iff $S$ is a smallest or largest subset of $\mathcal{E}_D$ with regards to set inclusion, respectively.
\end{definition}

One could observe the entire footprint of uttered arguments and generate an explanation by writing all their natural language representations, but this approach is too verbose and unwieldy in most cases (especially if agents operate under \textit{useful-single-argument} dialogue rules). We can attempt to specify a bound on the number of positions chosen to support an explanation.

\begin{definition}
We say $E' \in \mathcal{E}_D$ is an $n$\textit{-reason} explanation iff $|E'| = n$.
\end{definition}

We will now apply these definitions to do a proof of concept implementation using \acronym~ for the purposes of our user study.

\section{Proof of Concept Study}
\label{section:user-study}

In order to investigate the usefulness and efficiency of explanations in human-agent deconfliction settings, we designed a user study by instantiating a multi-agent resource contention environment. The problem of \textit{multi-agent path deconfliction} lends itself naturally to our objectives: it is a sufficiently intuitive problem, requires minimal prior knowledge, and disputed resources are obvious (space).

Our hypotheses are:

\begin{itemize}
    \item [$\mathbf{H_1}$:] Explanations provide a higher improvement for human performance in more complex systems than in simpler systems.
    \item [$\mathbf{H_2}$:] Explanations provide a higher decrease in time spent by a human in a task in more complex systems than in simpler systems.
    \item [$\mathbf{H_3}$:] User perception of explainable systems is more positive in more complex systems than in simpler systems.
\end{itemize}

Next, we introduce our definition of a path deconfliction environment and our application: the Busy Barracks game.

\subsection{Path Deconfliction Environment}

We define the path deconfliction environment in the form of a 2D discrete time and discrete space grid, represented by a DAG.

Let $L = (V,E)$ be a finite directed acyclic graph (DAG) whose vertices are contained within the points in $\mathbb{Z}^3$, representing a bi-dimensional discrete space as $x,y$ coordinates and time as $t$. Let $u = (x_1, y_1, t_1)$ and $v = (x_2, y_2, t_2)$ be any two points in this space. 

We denote $(u,v) \in E \iff (d(u,v) \leq d_{\text{max}}$ and $t_2 - t_1 = 1),$ where $d_{\text{max}}$ is the maximum distance achievable by any agent on a single time step, expressed as the Manhattan distance $d_M(u,v) = |x_1 - x_2| + |y_1 - y_2|$ between two points in a $G_{x \times y}$ grid graph.

 A set of $K$ obstacles $\Upsilon = \{\upsilon_1, \ldots, \upsilon_K\}$ is given as input, where $\Upsilon \subset V$. The resulting traversable graph $G$ is defined as $G = L - \Upsilon$.

A set of $N$ agents $\mathcal{Q} = \{q_1, \ldots, q_N\}$ is placed over $V(G)$. Each agent can traverse one edge for every time step. This edge may traverse longer distances in $(x,y)$ space, depending on the value of $d_{\text{max}}$ given as input. A goal $g_i \in V(G)$ is defined for every agent $q_i \in \mathcal{Q}$.

Plans to reach goal vertices are represented in the form of path subgraphs of $G$. Given an agent $q_i$ and its corresponding goal $g_i$, the agent's plan is represented in the form $P(q_i) = \{v_0, \ldots, v_i\}$, where $v_0$ is $q_i$'s current position and $v_i = g_i$. The length of plan $P(a_i)$ is equal to $|P(a_i)|$. 

\begin{definition}
\label{definition:conflicting-paths}
Two paths $P(a_i)$ and $P(a_j)$ are said to be \emph{conflicting} if $P(a_i) \cap P(a_j) \neq \emptyset$ (they attempt to visit the same vertex at the same time step) or if $\forall u = (x, y, t), \forall v = (x', y', t+1)$ s.t. $u, v \in P(a_i)$ and $\exists u' = (x', y', t), \exists v' = (x, y, t+1)$ s.t. $u', v' \in P(a_j)$, e.g., agents swap positions.
\end{definition}

\subsection{The Busy Barracks Game}

\begin{figure}
\includegraphics[width=8.5cm]{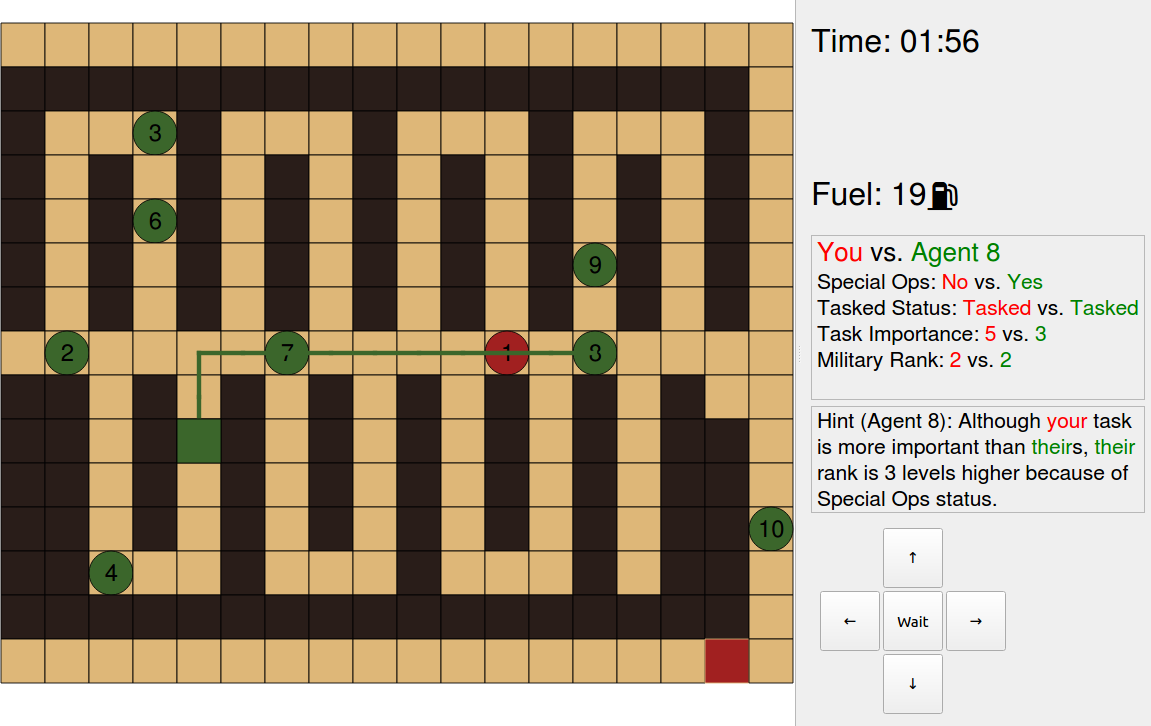}
\caption{The Busy Barracks game UI. Red player is human, followed by agents in green. The path to destination of Agent 8 can be seen as a green line. The human must decide whether to give way or to move towards the red square.}
\label{fig:gameui}
\end{figure}

We present the previously-defined Path Deconfliction Environment to human participants as a computer game called Busy Barracks. In the Busy Barracks (BB) game (see Figure~\ref{fig:gameui}), the human controls a military official represented by an agent $q_h \in \mathcal{Q}$. The human can choose one of two actions: move towards a direction (north, south, west, east), or choose to wait in place for a round. Agents move in lockstep, i.e., once the human makes a decision, all the agents make their planned move at the same time. The human is given 50 arbitrary units of fuel and told to navigate towards a goal destination under the following constraints:

\begin{itemize}
    \item For every move or wait action, the human will lose 1 unit of fuel.
    \item If a collision occurs (see Definition \ref{definition:conflicting-paths}), the human loses 5 units of fuel.
    \item Every 10 seconds past the first move in the game (in clock time), the human loses 1 unit of fuel.
\end{itemize}

The human is encouraged to reach their destination whilst maximising their remaining fuel. In practice, in order to achieve good scores, players will need to make short, collision-free trajectories with quick reaction times. The environment is composed of several other autonomous agents, who have individual goals and will also move towards their destinations concurrently with the player.

Humans are informed that agents will follow the rules and automatically reroute to clear the way if they understand that the human has priority in a specific conflict setting. Agents will remain in their original trajectory and expect the human to clear the way if they understand that they have priority according to the rules. It is down to the human to make the decision to either remain in their original trajectory (assuming that the agent will clear the way) or make way (assuming that the agent will keep their trajectory and will potentially collide if evasive action is not taken).

In order to explore the traits of a system with explicit rulesets, the human is provided with a deconfliction ruleset, presented in textual form on a sheet of paper. This document introduces arbitrary and game-specific properties that each agent has, and how those properties play out in generating a prioritisation when a spatial conflict arises. In other terms, by observing the properties and the rules correctly, every agent should unequivocally understand if they have the right of way or if they should concede and grant passage to the opponent.

\begin{example}
Suppose the following ruleset:
\begin{enumerate}
    \item You should have right of way if:
    \begin{enumerate}
        \item Your rank is higher than the other agent's rank.
        \item You are tasked and the other agent is not tasked, regardless of their rank.
    \end{enumerate}
\end{enumerate}

This ruleset implies the existence of two properties: \textit{rank} and \textit{tasked status}; and two rules: (a) and (b), as seen above. Thus, if we have agents $q_1: \{\text{rank}(q_1):2, \ \text{tasked}(q_1): yes\}$ and $q_2: \{\text{rank}(q_2): 4, \ \text{tasked}(q_2): no\}$, even though $q_2$ might be able to argue that it has a higher rank (rule (a)), it will be defeated when $q_1$ invokes rule (b).

Following this textual ruleset, we devise an example culture $C_{\textit{easy}} = (\mathcal{A}, \mathcal{R}, \mathcal{K})$. We instantiate the set of arguments $\mathcal{A} = \{\mu, a, b\}$, where $\mu$ represents the proposition (1) \textit{`you should have right of way'}. and $a, b$ represent rules (a) and (b), respectively. Let $c$ be a player and $\overline{c}$ their immediate opponent. The verifier functions are defined as follows:

\begin{equation}
\begin{aligned}
 f_{a}(c, \overline{c}) &= \begin{cases}
       \texttt{True} & \text{if rank}(c) > \text{rank}(\overline{c}),\\
       \texttt{False} & \text{otherwise.}
     \end{cases} 
\\
 f_{b}(c, \overline{c}) &= \begin{cases}
       \texttt{True} & \text{if tasked}(c) = \textit{yes} \ \text{and} \ \text{tasked}(\overline{c}) = \textit{no},\\
       \texttt{False} & \text{otherwise.}\\
     \end{cases}
\end{aligned}
\nonumber
\end{equation}

Since we know that rule (b) supersedes rule (a), we define $\mathcal{R} = \{(a, \mu), (b, \mu), (b, a)\}$ to complete the specification of $C_{\textit{easy}}$.
\label{example:culture-implementation}
\end{example}

\textbf{Culture: }For the BB game, we created three different rulesets, ranging in different levels of complexity. We posit that cultures become more complex as they grow in number of rules, hence our nomenclature. We refer back to the taxonomy seen in \citet{Rosenfeld2019ExplainabilitySystems} (\textit{not useful}, \textit{beneficial}, and \textit{critical}) to create three cultures with different sizes: \textit{easy}, \textit{medium}, and \textit{hard}. Each culture was created from a textual ruleset that was handed over to human players.

\begin{itemize}
    \item $C_{\textit{easy}}$: 2 properties and 2 rules (described in Example \ref{example:culture-implementation}.)
    \item $C_{\textit{medium}}$: 4 properties and 4 rules.
    \item $C_{\textit{hard}}$: 6 properties and 9 rules.
\end{itemize}

\textbf{Dialogue Game: }All players can publicly see the destination and intended trajectory of their opponents. When any two agents find themselves in conflict, they initiate a dialogue game and try to persuade the other to give way to them based on the culture that is being used in that instance of the game ($C_{\textit{easy}}, C_{\textit{medium}}, C_{\textit{hard}}.)$  In the BB game, all exchanges are \textit{useful-single-argument} dialogues. Moves are chosen randomly among the subset of demonstrably true arguments. The argumentative exchange happens in the background and is not visible to the human. 

The decision reached by this dialogue game decides the next action taken by the autonomous agent (to concede via rerouting or to continue in their original trajectory). The human must observe the rules and take action based on their belief of what the agent will do next. Agents always play optimally and do not make mistakes. A wrong decision from the human leads to two possible outcomes: either a collision or an unnecessary diversion from both human and agent, who both try to give way to each other (as the agent assumes the human will also play optimally.) There are 8 agents plus the human in every round, where exactly four of them will have right of way against the human, regardless of difficulty level. 

It is worth pointing out that the difficulty level does not affect the map layout, agent behaviour or any other factors that might influence scores or time other than the rules involved in deciding who gives way. If a human played with the same speed and the same success rate in every difficulty level, their scores would always be identical. Differences in score are uniquely defined by human performance.

\subsection{User Study}

We recruited 35 participants (21 male, 14 female, ages 20-39) within the university (students and staff). Participants were invited to play the BB game in a quiet room. Every new participant would be allocated to play one out of three versions of the game: either $C_{\textit{easy}}$, $C_{\textit{medium}}$, or $C_{\textit{hard}}$. Participants did not know that any other versions of the game were available.

\begin{figure}
\includegraphics[width=7.5cm]{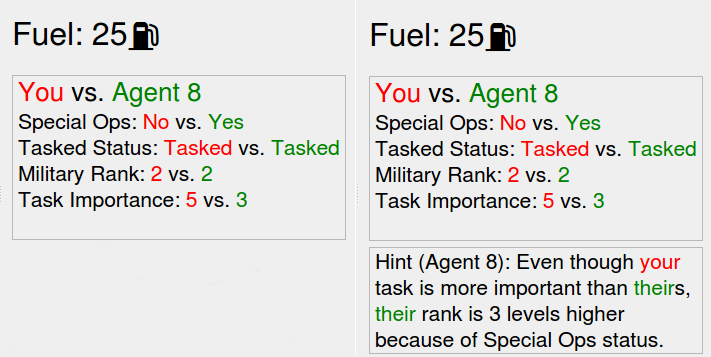}
\caption{Example of information available to human. Left: no explanation/hints, only data ($N$). Right: data + explanation/hint ($X$).}
\label{fig:hints}
\end{figure}

Our study is organised in a within-subject design in order to measure how each individual participant's performance is altered in the presence or absence of explanations. Each participant played two rounds of the game (within the same allocated culture): one version containing the only properties and rules, and another version containing properties, rules, and additional explanations generated in the form of \textit{hints} in the game UI (see Figure~\ref{fig:hints}). Those explanations were generated live based on the outcome of the background dialogue game between the human-controlled agent and the autonomous opponent. For brevity, we shall henceforth denote the \textit{non-explainable} round as $N$ and the \textit{explainable} round as $X$. We alternated the starting order of the rounds ($X$ or $N$) to minimise familiarity bias.

An experience questionnaire (extracted and modified from GEQ \cite{IJsselsteijn2013TheQuestionnaire}) was given to each participant at the end of each round. We clustered questions in three main groups (GEQ indices in brackets): Competence (10, 15, 17, 21); Affect (9, 22, 24); and Challenge (23, 26, 33). We included four custom questions to evaluate game-specific criteria, such as how often they consulted the text rules and if they anticipated/agreed with agents' actions. Answers were collected in a 5-point Likert scale. We collected game performance data, such as: score (represented by fuel units remaining at the end of the game), number of collisions, and time taken until completion.

The non-explainable version allows the human to visualise their opponent's trajectory and their properties. Based on this available information (and the rules' knowledge present in the ruleset), the human must then evaluate which rules apply and decide a course of action. In the explainable case, we decide to provide a succinct, or even partial explanation in the form of a hint. 

\textbf{Explanation Generation: }For that reason, in every dialogue game present in the game, \acronym~ generates hints by selecting a $2$\textit{-reason contrastive explanation} $CE'$ (a minimal and compact contrastive explanation) and presenting in a textual form as seen in Figure \ref{fig:hints}. Our objective is not to compare two versions with different information available -- but instead to evaluate the impact of having all the information required to make a decision ($N$) versus having all the information \textit{with the addition of} an explanation ($X$) to assist the human. 

We are interested in the differences in human performance between playing $N$ and $X$, namely, how much human performance improves or worsens between $N$ and $X$ in each difficulty level.

\section{Results}

Due to the limited number of samples, we choose to not make assumptions of parametrisation in the data. Every sample is grouped into a difficulty level ($E$, $M$, and $H$ representing $C_{\textit{easy}}$, $C_{\textit{medium}}$, and $C_{\textit{hard}}$, respectively). Users play two rounds ($N$ and $X$, in alternated order). Thus, a player allocated to $M$ would play both $MN$ and $MX$ rounds, respectively. Sample sizes are Easy (n = 11), Medium (n = 12), and Hard (n = 12).

We define our measures as:

\begin{itemize}
    \item Score ($S$): normalised score $(S_X - S_N)/(S_{\text{max}} - S_{\text{min}})$. Positive values of $S$ mean score improvement in $X$.
    \item Collisions ($Col$): normalised number of collisions $(Col_X - Col_N)/(Col_{\text{max}} - Col_{\text{min}})$. Negative values mean reduced number of collisions in $X$.
    \item Time ($T$): normalised time elapsed $(T_X - T_N)/(T_{\text{max}} - T_{\text{min}})$. Negative values of $T$ mean reduction in time elapsed in $X$.
\end{itemize}

\subsection{Score}

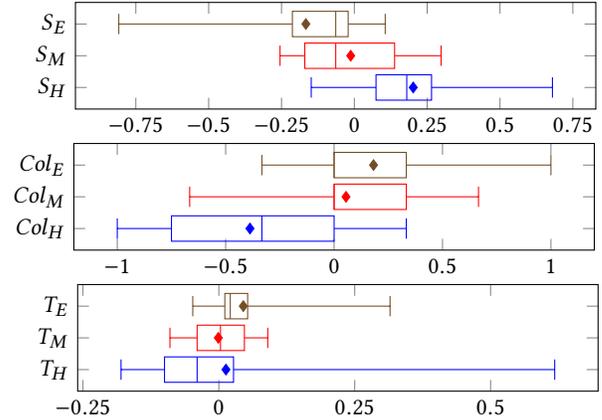
\begin{figure}
\begin{tikzpicture}
  \begin{axis}
    [
    ytick={1, 2, 3},
    yticklabels={$\quad \, S_H$, $\quad \, S_M$, $\quad \, S_E$},
    xtick={-1,-0.75,...,1},
    height=3cm, width=8.5cm,
    ]
    \addplot+[
    boxplot prepared={
      average=0.2021,
      upper whisker=0.680,
      upper quartile=0.265,
      median=0.180,
      lower quartile=0.0744,
      lower whisker=-0.148,
      sample size = 12
    },
    ] coordinates {};
    \addplot+[
    boxplot prepared={
      average=-0.0124,
      upper whisker=0.2978,
      upper quartile=0.1382,
      median=-0.0638,
      lower quartile=-0.170,
      lower whisker=-0.255,
      sample size = 12
    },
    ] coordinates {};
     \addplot+[
    boxplot prepared={
      average=-0.1663,
      upper whisker=0.106,
      upper quartile=-0.0212,
      median=-0.0638,
      lower quartile=-0.2127,
      lower whisker=-0.8085,
      sample size = 11
    },
    ] coordinates {};

  \end{axis}
\end{tikzpicture}
\begin{tikzpicture}
  \begin{axis}
    [
    ytick={1, 2, 3},
    yticklabels={$Col_H$, $Col_M$, $Col_E$},
    xtick={-1,-0.5,...,1},
    height=3cm, width=8.5cm,
    ]
    \addplot+[
    boxplot prepared={
      average=-0.388,
      upper whisker=0.333,
      upper quartile=0,
      median=-0.333,
      lower quartile=-0.75,
      lower whisker=-1,
      sample size = 12
    },
    ] coordinates {};
    \addplot+[
    boxplot prepared={
      average=0.055,
      upper whisker=0.666,
      upper quartile=0.333,
      median=0,
      lower quartile=0,
      lower whisker=-0.666,
      sample size = 12
    },
    ] coordinates {};
     \addplot+[
    boxplot prepared={
      average=0.1818,
      upper whisker=1,
      upper quartile=0.333,
      median=0,
      lower quartile=0,
      lower whisker=-0.333,
      sample size = 11
    },
    ] coordinates {};
  \end{axis}
\end{tikzpicture}
\begin{tikzpicture}
  \begin{axis}
    [
    ytick={1, 2, 3},
    yticklabels={$\quad \, T_H$, $\quad \, T_M$, $\quad \, T_E$},
    xtick={-1, -0.75,..., 1},
    height=3cm, width=8.5cm,
    ]
    \addplot+[
    boxplot prepared={
      average=0.013,
      upper whisker=0.618,
      upper quartile=0.027,
      median=-0.04,
      lower quartile=-0.10,
      lower whisker=-0.18,
      sample size = 12
    },
    ] coordinates {};
    \addplot+[
    boxplot prepared={
      average=-0.001,
      upper whisker=0.09,
      upper quartile=0.047,
      median=0.003,
      lower quartile=-0.04,
      lower whisker=-0.09,
      sample size = 12
    },
    ] coordinates {};
     \addplot+[
    boxplot prepared={
      average=0.045,
      upper whisker=0.315,
      upper quartile=0.053,
      median=0.021,
      lower quartile=0.011,
      lower whisker=-0.048,
      sample size = 11
    },
    ] coordinates {};
  \end{axis}
\end{tikzpicture}
\caption{Scores ($S$), Collisions ($Col$), and Time ($T$). Results are shown in the form of box plots (25th, 50th, 75th percentile, and whiskers covering all data and outliers).}
\label{fig:deltascore}
\end{figure}

Given 3 sample sets: $S_E$ ($E$ scores), $S_M$ ($M$ scores), and $S_H$ ($H$ scores) (see Figure \ref{fig:deltascore}), we run a Kruskal-Wallis H-Test (KW) under the alternative hypothesis that at least one of the distributions come from a different population and confirm significant differences (H = 11.63, p = 0.003**)\footnote{(*) p < 0.05; (**) p < 0.01; (***) p < 0.001.}. We then perform a pairwise one-sided Mann-Whitney U-Test (MW) under the alternative hypothesis that easier categories have significantly smaller $S$ than harder ones, meaning that the improvement in $X$ is less pronounced in easier rounds.

\begin{table}[]
\centering
\begin{tabular}{lllll}
\cline{2-4}
\multicolumn{1}{p{0.75cm}|}{}     & \multicolumn{1}{l|}{$S_E$} & \multicolumn{1}{l|}{$S_M$}                                                           & \multicolumn{1}{p{1.5cm}|}{$S_H$}                                                              &  \\ \cline{1-4}
\multicolumn{1}{|l|}{$S_E$} & \multicolumn{1}{p{1.25cm}|}{-}   & \multicolumn{1}{p{1.25cm}|}{\begin{tabular}[c]{@{}l@{}}U = 48.5\\ p = 0.1472\end{tabular}} & \multicolumn{1}{p{1.25cm}|}{\begin{tabular}[c]{@{}l@{}}U = 11.5\\ p = 0.0004***\end{tabular}} &  \\ \cline{1-4}
\multicolumn{1}{|l|}{$S_M$} & \multicolumn{1}{p{1.25cm}|}{-}   & \multicolumn{1}{l|}{-}                                                             & \multicolumn{1}{p{1.25cm}|}{\begin{tabular}[c]{@{}l@{}}U = 34.0\\ p = 0.015*\end{tabular}}    &  \\ \cline{1-4}
                          &                          &                                                                                    &                                                                                       & 
\end{tabular}
\begin{tabular}{lllll}
\cline{2-4}
\multicolumn{1}{p{0.75cm}|}{}     & \multicolumn{1}{l|}{$Col_E$} & \multicolumn{1}{l|}{$Col_M$}                                                          & \multicolumn{1}{p{1.5cm}|}{$Col_H$}                                                             &  \\ \cline{1-4}
\multicolumn{1}{|l|}{$Col_E$} & \multicolumn{1}{p{1.25cm}|}{-}   & \multicolumn{1}{p{1.25cm}|}{\begin{tabular}[c]{@{}l@{}}U = 72.5\\ p = 0.346\end{tabular}} & \multicolumn{1}{p{1.25cm}|}{\begin{tabular}[c]{@{}l@{}}U = 111.5\\ p = 0.002**\end{tabular}} &  \\ \cline{1-4}
\multicolumn{1}{|l|}{$Col_M$} & \multicolumn{1}{p{1.25cm}|}{-}   & \multicolumn{1}{l|}{-}                                                            & \multicolumn{1}{p{1.25cm}|}{\begin{tabular}[c]{@{}l@{}}U = 112.5\\ p = 0.008**\end{tabular}} &  \\ \cline{1-4}
                          &                          &                                                                                   &                                                                                      & 
\end{tabular}
\begin{tabular}{lllll}
\cline{2-4}
\multicolumn{1}{l|}{}     & \multicolumn{1}{l|}{$Cha_E$} & \multicolumn{1}{l|}{$Cha_M$}                                                          & \multicolumn{1}{p{1.5cm}|}{$Cha_H$}                                                             &  \\ \cline{1-4}
\multicolumn{1}{|l|}{$Cha_E$} & \multicolumn{1}{p{1.25cm}|}{-}   & \multicolumn{1}{p{1.25cm}|}{\begin{tabular}[c]{@{}l@{}}U = 92.0\\ p = 0.056\end{tabular}} & \multicolumn{1}{p{1.25cm}|}{\begin{tabular}[c]{@{}l@{}}U = 116.5\\ p = 0.001***\end{tabular}} &  \\ \cline{1-4}
\multicolumn{1}{|l|}{$Cha_M$} & \multicolumn{1}{p{1.25cm}|}{-}   & \multicolumn{1}{l|}{-}                                                            & \multicolumn{1}{p{1.25cm}|}{\begin{tabular}[c]{@{}l@{}}U = 101.0\\ p = 0.047*\end{tabular}} &  \\ \cline{1-4}
                          &                          &                                                                                   &                                                                                      & 
\end{tabular}
\caption{Pairwise MW for $S$, $Col$, and $Cha$.}
\label{table:mwu-ds}
\end{table}

The results in Table \ref{table:mwu-ds} show that score improvement in $M$ is significantly smaller than $H$, whilst score improvement in $E$ is \textit{very} significantly smaller than $H$, although not significantly smaller than $M$.

\subsection{Collisions}

Like the previous sets, we consider $Col$ to isolate the number of wrong decisions that specifically led to collisions (see Definition \ref{definition:conflicting-paths}), and how did that differ within subjects between $N$ and $X$. We run a KW under the alternative hypothesis that at least one of the distributions come from a different population, confirming the hypothesis (H = 9.83, p = 0.007**). 

Since the distributions are different, we perform a one-sided MW, this time with the alternative hypothesis that easier categories have significantly \textit{higher} number of collisions, i.e., they do not improve (and reduce) their number of collisions as well as harder levels. The results in Table \ref{table:mwu-ds} show that collision improvement in $M$ is significantly smaller than $H$, whilst collision improvement in $E$ is also significantly smaller than $H$, although not significantly smaller than $M$. Both $S$ and $Col$ results support $\mathbf{H_1}$.

\subsection{Times}

Similarly to $Col$, $T$ represents the change in time elapsed to complete each round from $N$ to $X$. We run a KW in order to isolate the distributions but did not find significant differences (H = 3.61, p = 0.16). However, a pairwise one-sided MW reveals a significant improvement in $T$ between $E$ and $H$ (U = 94.0, p = 0.04*), showing that participants in $H$ have a superior reduction in time in $X$ compared to those in $E$. This result supports $\mathbf{H_2}$.

\subsection{User Experience}

In order to evaluate the effect of ordering (whether users who played their first round as $N$ or $X$ had a significantly different perception of the game), we ran KWs for each cluster of questions, separating populations by their starting mode ($N$ or $X$). The alternative hypothesis for all cases was that there was a significant difference in populations, which was not confirmed for any: Challenge (H = 0.24, p = 0.61); Competence (H = 0.48, p = 0.48); Affect (H = 0.13, p = 0.71); and Game-Specific (H = 1.73, p = 0.18).

We then evaluate the populations based on the difficulty level. To evaluate the differences in populations, we ran KWs under the alternative hypothesis that the populations differ significantly depending on difficulty level. We manage to validate this hypothesis for Challenge ($Cha$) (H = 10.28, p = 0.005**), but not for Affect ($At$) (H = 3.91, p = 0.14), Competence ($Com$) (H = 4.98, p = 0.08) and Game-Specific ($Gam$) (H = 5.52, p = 0.06). We perform pairwise one-sided MWs under the alternative hypothesis that easier categories have a significantly smaller improvement in the perception of challenge from $N$ to $X$.

The results in Table~\ref{table:mwu-ds} show that the improvement of perception of $Cha$ in $M$ is significantly smaller than $H$. The improvement in $E$ is \textit{very} significantly smaller than $H$, although not significantly smaller than $M$. Despite populations being not isolated in the previous KW for $Gam$, similar pairwise MW results are found: $Gam_E$ vs.~ $Gam_H$ (U = 56.5, p = 0.01*) and $Gam_M$ vs.~ $Gam_H$ (U = 42.5, p = 0.04*).

Additionally, user experience results in $At$ and $Com$ clusters also revealed significant improvements ($At$: U = 96.0, p = 0.03*; $Com$: U = 32.0, p = 0.02*) from $N$ to $X$ between between $E$ and $H$ levels. These results support $\mathbf{H_3}$.

\vspace{-0.5em}

\section{Discussion and Conclusion}

We achieved significant results in demonstrating how the benefit of explanations in human-agent deconfliction correlates to the complexity of the underlying system. Our results demonstrated clear differences between within-subject improvement when comparing their performance in $N$ against the performance in $X$, which demonstrates that humans benefit from explanations -- but mostly when the system is sufficiently complex to warrant such explanations.

In fact, when the complexity is small, humans might actually perform better \textit{without} any explanations. We probe this claim by running a one-sided MW considering the alternative hypothesis that global (between-subjects) $EN$ scores were higher than $EX$ scores (U = 88.0, p = 0.03*), which was significant. Contrariwise, a similar test under the alternative hypothesis that global $HN$ scores are \textit{lower} than $HX$ scores (U = 36.0, p = 0.019*) also proved significant. In many cases, $M$ populations were harder to distinguish between $E$ and $H$ in nondirectional tests, such as in $T$, $At$, and $Com$ analyses. Still, hypotheses $\mathbf{H_1}$, $\mathbf{H_2}$, and $\mathbf{H_3}$ are validated for all $E$ and $H$ within-subject results. We believe that a larger scale study and further refinement of $M$ in terms of complexity might consolidate all populations more clearly.

Post-experiment interviews were conducted to discuss the user experience. Participants were asked to self-report on how they felt about the hints. Six out of 11 participants who played the $E$ version reported finding the hints \textit{not useful}. At the $M$ level (n = 12), 4 participants found the hints \textit{not useful}, and 5 expressed using hints as a useful confirmation mechanism to check their mental computation. Last, at $H$ (n = 12), 9 players reported that hints were \textit{very useful} and \textit{primarily relied} on the hints to act. These findings map well to the taxonomy of \cite{Rosenfeld2019ExplainabilitySystems} (\textit{not useful}, \textit{beneficial}, and \textit{critical}) and suggest that the taxonomy of the \textit{need} for explanations can be considered under a new dimension: that of system complexity.

In this game, the deconfliction behaviour of the agents was entirely dictated by the present culture. We believe that \acronym~ can find applications beyond human-agent deconfliction, and towards multi-agent systems in general. For example, a decentralised multi-agent system could be designed in terms of a culture, and performing individual implementations for each rule could prove easier than writing a monolithic policy that tries to emulate a complex ruleset (especially if coming from text/human regulations), with the innate benefit of being explainable, as demonstrated by our architecture and study. The deconfliction behaviour of agents can be changed by adding or removing arguments individually, or changing their attack relationships. Future studies will demonstrate how \acronym~ can be used for modelling real-life rulesets in deployed multi-agent/multi-robot applications.

\begin{acks}
  The authors would like to thank Dr Jon Roozenbeek, Nikhil Churamani, and Guilherme Paulino-Passos for providing valuable assistance during this study. Alex Raymond is supported by L3Harris ASV and the Royal Commission for the Exhibition of 1851. Hatice Gunes is supported by the EPSRC Project ARoEQ (Grant Ref: EP/R030782/1). Amanda Prorok is supported by the EPSRC (Grant Ref: EP/S015493/1). Their support is gratefully acknowledged. 

\end{acks}



\begin{thebibliography}{00}

%%% the Web address.

\ifx \showCODEN    \undefined \def \showCODEN     #1{\unskip}     \fi
\ifx \showDOI      \undefined \def \showDOI       #1{#1}\fi
\ifx \showISBNx    \undefined \def \showISBNx     #1{\unskip}     \fi
\ifx \showISBNxiii \undefined \def \showISBNxiii  #1{\unskip}     \fi
\ifx \showISSN     \undefined \def \showISSN      #1{\unskip}     \fi
\ifx \showLCCN     \undefined \def \showLCCN      #1{\unskip}     \fi
\ifx \shownote     \undefined \def \shownote      #1{#1}          \fi
\ifx \showarticletitle \undefined \def \showarticletitle #1{#1}   \fi
\ifx \showURL      \undefined \def \showURL       {\relax}        \fi
% invisible to TeX
\providecommand\bibfield[2]{#2}
\providecommand\bibinfo[2]{#2}
\providecommand\natexlab[1]{#1}
\providecommand\showeprint[2][]{arXiv:#2}

\bibitem[\protect\citeauthoryear{Amgoud and Prade}{Amgoud and Prade}{2009}]%
        {Amgoud2009UsingDecisions}
\bibfield{author}{\bibinfo{person}{Leila Amgoud} {and} \bibinfo{person}{Henri
  Prade}.} \bibinfo{year}{2009}\natexlab{}.
\newblock \showarticletitle{{Using arguments for making and explaining
  decisions}}.
\newblock \bibinfo{journal}{{\em Artificial Intelligence\/}}
  \bibinfo{volume}{173}, \bibinfo{number}{3-4} (\bibinfo{year}{2009}),
  \bibinfo{pages}{413--436}.
\newblock


\bibitem[\protect\citeauthoryear{Anjomshoae, Najjar, Calvaresi, and
  Fr{\"{a}}mling}{Anjomshoae et~al\mbox{.}}{2019}]%
        {Anjomshoae2019ExplainableReview}
\bibfield{author}{\bibinfo{person}{Sule Anjomshoae}, \bibinfo{person}{Amro
  Najjar}, \bibinfo{person}{Davide Calvaresi}, {and} \bibinfo{person}{Kary
  Fr{\"{a}}mling}.} \bibinfo{year}{2019}\natexlab{}.
\newblock \showarticletitle{{Explainable Agents and Robots: Results from a
  Systematic Literature Review}}. In \bibinfo{booktitle}{{\em Proceedings of
  the 18th International Conference on Autonomous Agents and MultiAgent
  Systems}} {\em (\bibinfo{series}{AAMAS '19})}.
  \bibinfo{publisher}{International Foundation for Autonomous Agents and
  Multiagent Systems}, \bibinfo{address}{Richland, SC},
  \bibinfo{pages}{1078--1088}.
\newblock
\showISBNx{978-1-4503-6309-9}
\showURL{%
\url{http://dl.acm.org/citation.cfm?id=3306127.3331806}}


\bibitem[\protect\citeauthoryear{Asaro}{Asaro}{2016}]%
        {Asaro2016TheAgents}
\bibfield{author}{\bibinfo{person}{Peter~M Asaro}.}
  \bibinfo{year}{2016}\natexlab{}.
\newblock \showarticletitle{{The liability problem for autonomous artificial
  agents}}. In \bibinfo{booktitle}{{\em 2016 AAAI Spring Symposium Series}}.
\newblock


\bibitem[\protect\citeauthoryear{Bench-Capon and Modgil}{Bench-Capon and
  Modgil}{2019}]%
        {Bench-Capon2019NormsFrameworks}
\bibfield{author}{\bibinfo{person}{Trevor Bench-Capon} {and}
  \bibinfo{person}{Sanjay Modgil}.} \bibinfo{year}{2019}\natexlab{}.
\newblock \showarticletitle{{Norms and Extended Argumentation Frameworks}}. In
  \bibinfo{booktitle}{{\em Proceedings of the Seventeenth International
  Conference on Artificial Intelligence and Law}} {\em (\bibinfo{series}{ICAIL
  '19})}. \bibinfo{publisher}{ACM}, \bibinfo{address}{New York, NY, USA},
  \bibinfo{pages}{174--178}.
\newblock
\showISBNx{978-1-4503-6754-7}
\showDOI{%
\url{https://doi.org/10.1145/3322640.3326696}}


\bibitem[\protect\citeauthoryear{Benjamin, Curcio, Leonard, and
  Newman}{Benjamin et~al\mbox{.}}{2006}]%
        {Benjamin2006NavigationRoad}
\bibfield{author}{\bibinfo{person}{Michael~R Benjamin},
  \bibinfo{person}{Joseph~A Curcio}, \bibinfo{person}{John~J Leonard}, {and}
  \bibinfo{person}{Paul~M Newman}.} \bibinfo{year}{2006}\natexlab{}.
\newblock \showarticletitle{{Navigation of unmanned marine vehicles in
  accordance with the rules of the road}}. In \bibinfo{booktitle}{{\em
  Proceedings 2006 IEEE International Conference on Robotics and Automation,
  2006. ICRA 2006.}} \bibinfo{pages}{3581--3587}.
\newblock


\bibitem[\protect\citeauthoryear{Bentahar, Moulin, and B{\'{e}}langer}{Bentahar
  et~al\mbox{.}}{2010}]%
        {Bentahar2010ARepresentation}
\bibfield{author}{\bibinfo{person}{Jamal Bentahar}, \bibinfo{person}{Bernard
  Moulin}, {and} \bibinfo{person}{Micheline B{\'{e}}langer}.}
  \bibinfo{year}{2010}\natexlab{}.
\newblock \showarticletitle{{A taxonomy of argumentation models used for
  knowledge representation}}.
\newblock \bibinfo{journal}{{\em Artificial Intelligence Review\/}}
  \bibinfo{volume}{33}, \bibinfo{number}{3} (\bibinfo{date}{3}
  \bibinfo{year}{2010}), \bibinfo{pages}{211--259}.
\newblock
\showISSN{0269-2821}
\showDOI{%
\url{https://doi.org/10.1007/s10462-010-9154-1}}


\bibitem[\protect\citeauthoryear{Billhardt, Juli{\'{a}}n, Corchado, and
  Fern{\'{a}}ndez}{Billhardt et~al\mbox{.}}{2014}]%
        {Billhardt2014AnSocieties}
\bibfield{author}{\bibinfo{person}{Holger Billhardt}, \bibinfo{person}{Vicente
  Juli{\'{a}}n}, \bibinfo{person}{Juan~Manuel Corchado}, {and}
  \bibinfo{person}{Alberto Fern{\'{a}}ndez}.} \bibinfo{year}{2014}\natexlab{}.
\newblock \showarticletitle{{An architecture proposal for human-agent
  societies}}. In \bibinfo{booktitle}{{\em International Conference on
  Practical Applications of Agents and Multi-Agent Systems}}.
  \bibinfo{pages}{344--357}.
\newblock


\bibitem[\protect\citeauthoryear{Coste-Marquis, Devred, and
  Marquis}{Coste-Marquis et~al\mbox{.}}{2005}]%
        {Coste-Marquis2005SymmetricFrameworks}
\bibfield{author}{\bibinfo{person}{Sylvie Coste-Marquis},
  \bibinfo{person}{Caroline Devred}, {and} \bibinfo{person}{Pierre Marquis}.}
  \bibinfo{year}{2005}\natexlab{}.
\newblock \showarticletitle{{Symmetric Argumentation Frameworks}}.
\newblock In \bibinfo{booktitle}{{\em Symbolic and Quantitative Approaches to
  Reasoning with Uncertainty}}. \bibinfo{publisher}{Springer, Berlin,
  Heidelberg}, \bibinfo{pages}{317--328}.
\newblock
\showDOI{%
\url{https://doi.org/10.1007/11518655{\_}28}}


\bibitem[\protect\citeauthoryear{Cranefield, Oren, and Vasconcelos}{Cranefield
  et~al\mbox{.}}{2019}]%
        {Cranefield2019AccountabilityAgents}
\bibfield{author}{\bibinfo{person}{Stephen Cranefield}, \bibinfo{person}{Nir
  Oren}, {and} \bibinfo{person}{Wamberto~W Vasconcelos}.}
  \bibinfo{year}{2019}\natexlab{}.
\newblock \showarticletitle{{Accountability for Practical Reasoning Agents}}.
  In \bibinfo{booktitle}{{\em Agreement Technologies}},
  \bibfield{editor}{\bibinfo{person}{Marin Lujak}} (Ed.).
  \bibinfo{publisher}{Springer International Publishing},
  \bibinfo{address}{Cham}, \bibinfo{pages}{33--48}.
\newblock
\showISBNx{978-3-030-17294-7}


\bibitem[\protect\citeauthoryear{Dung}{Dung}{1995}]%
        {Dung1995OnGames}
\bibfield{author}{\bibinfo{person}{Phan~Minh Dung}.}
  \bibinfo{year}{1995}\natexlab{}.
\newblock \showarticletitle{{On the acceptability of arguments and its
  fundamental role in nonmonotonic reasoning, logic programming and n-person
  games}}.
\newblock \bibinfo{journal}{{\em Artificial Intelligence\/}}
  \bibinfo{volume}{77}, \bibinfo{number}{2} (\bibinfo{date}{9}
  \bibinfo{year}{1995}), \bibinfo{pages}{321--357}.
\newblock
\showISSN{0004-3702}
\showDOI{%
\url{https://doi.org/10.1016/0004-3702(94)00041-X}}


\bibitem[\protect\citeauthoryear{Fan and Toni}{Fan and Toni}{2015}]%
        {Fan2015OnArgumentation}
\bibfield{author}{\bibinfo{person}{Xiuyi Fan} {and} \bibinfo{person}{Francesca
  Toni}.} \bibinfo{year}{2015}\natexlab{}.
\newblock \showarticletitle{{On Computing Explanations in Argumentation}}.
\newblock \bibinfo{journal}{{\em Twenty-Ninth AAAI Conference on Artificial
  Intelligence\/}} (\bibinfo{date}{2} \bibinfo{year}{2015}).
\newblock
\showURL{%
\url{https://www.aaai.org/ocs/index.php/AAAI/AAAI15/paper/viewPaper/9872}}


\bibitem[\protect\citeauthoryear{Finkelstein}{Finkelstein}{1968}]%
        {Finkelstein1968TheUr-Nammu}
\bibfield{author}{\bibinfo{person}{Jacob~J Finkelstein}.}
  \bibinfo{year}{1968}\natexlab{}.
\newblock \showarticletitle{{The Laws of Ur-Nammu}}.
\newblock \bibinfo{journal}{{\em Journal of cuneiform studies\/}}
  \bibinfo{volume}{22}, \bibinfo{number}{3-4} (\bibinfo{year}{1968}),
  \bibinfo{pages}{66--82}.
\newblock


\bibitem[\protect\citeauthoryear{Gunning}{Gunning}{2017}]%
        {Gunning2017Explainablexai}
\bibfield{author}{\bibinfo{person}{David Gunning}.}
  \bibinfo{year}{2017}\natexlab{}.
\newblock \showarticletitle{{Explainable artificial intelligence (xai)}}.
\newblock \bibinfo{journal}{{\em Defense Advanced Research Projects Agency
  (DARPA), nd Web\/}}  \bibinfo{volume}{2} (\bibinfo{year}{2017}).
\newblock


\bibitem[\protect\citeauthoryear{IJsselsteijn, De~Kort, and Poels}{IJsselsteijn
  et~al\mbox{.}}{2013}]%
        {IJsselsteijn2013TheQuestionnaire}
\bibfield{author}{\bibinfo{person}{W~A IJsselsteijn}, \bibinfo{person}{Y~A~W
  De~Kort}, {and} \bibinfo{person}{Karolien Poels}.}
  \bibinfo{year}{2013}\natexlab{}.
\newblock \showarticletitle{{The game experience questionnaire}}.
\newblock \bibinfo{journal}{{\em Eindhoven: Technische Universiteit
  Eindhoven\/}} (\bibinfo{year}{2013}).
\newblock


\bibitem[\protect\citeauthoryear{Jakobovits and Vermeir}{Jakobovits and
  Vermeir}{1999}]%
        {Jakobovits1999DialecticFrameworks}
\bibfield{author}{\bibinfo{person}{H. Jakobovits} {and} \bibinfo{person}{D.
  Vermeir}.} \bibinfo{year}{1999}\natexlab{}.
\newblock \showarticletitle{{Dialectic semantics for argumentation
  frameworks}}. In \bibinfo{booktitle}{{\em Proceedings of the seventh
  international conference on Artificial intelligence and law - ICAIL '99}}.
  \bibinfo{publisher}{ACM Press}, \bibinfo{address}{New York, New York, USA},
  \bibinfo{pages}{53--62}.
\newblock
\showISBNx{1581131658}
\showDOI{%
\url{https://doi.org/10.1145/323706.323715}}


\bibitem[\protect\citeauthoryear{Kakas and Moraitis}{Kakas and
  Moraitis}{2003}]%
        {Kakas2003ArgumentationAgents}
\bibfield{author}{\bibinfo{person}{Antonis Kakas} {and} \bibinfo{person}{Pavlos
  Moraitis}.} \bibinfo{year}{2003}\natexlab{}.
\newblock \showarticletitle{{Argumentation Based Decision Making for Autonomous
  Agents}}. In \bibinfo{booktitle}{{\em Proceedings of the Second International
  Joint Conference on Autonomous Agents and Multiagent Systems}} {\em
  (\bibinfo{series}{AAMAS '03})}. \bibinfo{publisher}{ACM},
  \bibinfo{address}{New York, NY, USA}, \bibinfo{pages}{883--890}.
\newblock
\showISBNx{1-58113-683-8}
\showDOI{%
\url{https://doi.org/10.1145/860575.860717}}


\bibitem[\protect\citeauthoryear{Karamlou, {\v{C}}yras, and Toni}{Karamlou
  et~al\mbox{.}}{2019}]%
        {Karamlou2019DecidingArgumentation}
\bibfield{author}{\bibinfo{person}{Amin Karamlou}, \bibinfo{person}{Kristijonas
  {\v{C}}yras}, {and} \bibinfo{person}{Francesca Toni}.}
  \bibinfo{year}{2019}\natexlab{}.
\newblock \showarticletitle{{Deciding the Winner of a Debate Using Bipolar
  Argumentation}}. In \bibinfo{booktitle}{{\em Proceedings of the 18th
  International Conference on Autonomous Agents and MultiAgent Systems}} {\em
  (\bibinfo{series}{AAMAS '19})}. \bibinfo{publisher}{International Foundation
  for Autonomous Agents and Multiagent Systems}, \bibinfo{address}{Richland,
  SC}, \bibinfo{pages}{2366--2368}.
\newblock
\showISBNx{978-1-4503-6309-9}
\showURL{%
\url{http://dl.acm.org/citation.cfm?id=3306127.3332114}}


\bibitem[\protect\citeauthoryear{Langley, Meadows, Sridharan, and Choi}{Langley
  et~al\mbox{.}}{2017}]%
        {Langley2017ExplainableSystems}
\bibfield{author}{\bibinfo{person}{Pat Langley}, \bibinfo{person}{Ben Meadows},
  \bibinfo{person}{Mohan Sridharan}, {and} \bibinfo{person}{Dongkyu Choi}.}
  \bibinfo{year}{2017}\natexlab{}.
\newblock \showarticletitle{{Explainable Agency for Intelligent Autonomous
  Systems}}.
\newblock \bibinfo{journal}{{\em Twenty-Ninth IAAI Conference\/}}
  (\bibinfo{date}{2} \bibinfo{year}{2017}).
\newblock
\showURL{%
\url{https://www.aaai.org/ocs/index.php/IAAI/IAAI17/paper/viewPaper/15046}}


\bibitem[\protect\citeauthoryear{Modgil}{Modgil}{2014}]%
        {Modgil2014RevisitingFrameworks}
\bibfield{author}{\bibinfo{person}{Sanjay Modgil}.}
  \bibinfo{year}{2014}\natexlab{}.
\newblock \showarticletitle{{Revisiting Abstract Argumentation Frameworks}}.
\newblock In \bibinfo{booktitle}{{\em Theory and Applications of Formal
  Argumentation}}. \bibinfo{publisher}{Springer, Berlin, Heidelberg},
  \bibinfo{pages}{1--15}.
\newblock
\showDOI{%
\url{https://doi.org/10.1007/978-3-642-54373-9{\_}1}}


\bibitem[\protect\citeauthoryear{Modgil and Caminada}{Modgil and
  Caminada}{2009}]%
        {Modgil2009ProofFrameworks}
\bibfield{author}{\bibinfo{person}{Sanjay Modgil} {and} \bibinfo{person}{Martin
  Caminada}.} \bibinfo{year}{2009}\natexlab{}.
\newblock \showarticletitle{{Proof Theories and Algorithms for Abstract
  Argumentation Frameworks}}.
\newblock In \bibinfo{booktitle}{{\em Argumentation in Artificial
  Intelligence}}. \bibinfo{publisher}{Springer US}, \bibinfo{address}{Boston,
  MA}, \bibinfo{pages}{105--129}.
\newblock
\showDOI{%
\url{https://doi.org/10.1007/978-0-387-98197-0{\_}6}}


\bibitem[\protect\citeauthoryear{Pagallo}{Pagallo}{2016}]%
        {Pagallo2016EvenLaw}
\bibfield{author}{\bibinfo{person}{Ugo Pagallo}.}
  \bibinfo{year}{2016}\natexlab{}.
\newblock \showarticletitle{{Even Angels Need the Rules: AI, Roboethics, and
  the Law}}. In \bibinfo{booktitle}{{\em Proceedings of the Twenty-second
  European Conference on Artificial Intelligence}} {\em
  (\bibinfo{series}{ECAI'16})}. \bibinfo{publisher}{IOS Press},
  \bibinfo{address}{Amsterdam, The Netherlands, The Netherlands},
  \bibinfo{pages}{209--215}.
\newblock
\showISBNx{978-1-61499-671-2}
\showDOI{%
\url{https://doi.org/10.3233/978-1-61499-672-9-209}}


\bibitem[\protect\citeauthoryear{Rizaldi and Althoff}{Rizaldi and
  Althoff}{2015}]%
        {Rizaldi2015FormalisingVehicles}
\bibfield{author}{\bibinfo{person}{Albert Rizaldi} {and}
  \bibinfo{person}{Matthias Althoff}.} \bibinfo{year}{2015}\natexlab{}.
\newblock \showarticletitle{{Formalising traffic rules for accountability of
  autonomous vehicles}}. In \bibinfo{booktitle}{{\em 2015 IEEE 18th
  International Conference on Intelligent Transportation Systems}}.
  \bibinfo{pages}{1658--1665}.
\newblock


\bibitem[\protect\citeauthoryear{Rosenfeld and Kraus}{Rosenfeld and
  Kraus}{2016}]%
        {Rosenfeld2016StrategicalPersuasion}
\bibfield{author}{\bibinfo{person}{Ariel Rosenfeld} {and}
  \bibinfo{person}{Sarit Kraus}.} \bibinfo{year}{2016}\natexlab{}.
\newblock \showarticletitle{{Strategical argumentative agent for human
  persuasion}}. In \bibinfo{booktitle}{{\em Proceedings of the Twenty-second
  European Conference on Artificial Intelligence}}. \bibinfo{pages}{320--328}.
\newblock


\bibitem[\protect\citeauthoryear{Rosenfeld and Richardson}{Rosenfeld and
  Richardson}{2019}]%
        {Rosenfeld2019ExplainabilitySystems}
\bibfield{author}{\bibinfo{person}{Avi Rosenfeld} {and}
  \bibinfo{person}{Ariella Richardson}.} \bibinfo{year}{2019}\natexlab{}.
\newblock \showarticletitle{{Explainability in human?agent systems}}.
\newblock \bibinfo{journal}{{\em Autonomous Agents and Multi-Agent Systems\/}}
  \bibinfo{volume}{33}, \bibinfo{number}{6} (\bibinfo{date}{11}
  \bibinfo{year}{2019}), \bibinfo{pages}{673--705}.
\newblock
\showDOI{%
\url{https://doi.org/10.1007/s10458-019-09408-y}}


\bibitem[\protect\citeauthoryear{Sklar and Azhar}{Sklar and Azhar}{2018}]%
        {Sklar2018ExplanationArgumentation}
\bibfield{author}{\bibinfo{person}{Elizabeth~I. Sklar} {and}
  \bibinfo{person}{Mohammad~Q. Azhar}.} \bibinfo{year}{2018}\natexlab{}.
\newblock \showarticletitle{{Explanation through Argumentation}}. In
  \bibinfo{booktitle}{{\em Proceedings of the 6th International Conference on
  Human-Agent Interaction - HAI '18}}. \bibinfo{publisher}{ACM Press},
  \bibinfo{address}{New York, New York, USA}, \bibinfo{pages}{277--285}.
\newblock
\showISBNx{9781450359535}
\showDOI{%
\url{https://doi.org/10.1145/3284432.3284470}}


\bibitem[\protect\citeauthoryear{Young, K{\"{o}}kciyan, Sassoon, Modgil, and
  Parsons}{Young et~al\mbox{.}}{2018}]%
        {Young2018InstantiatingFrameworks}
\bibfield{author}{\bibinfo{person}{Anthony~P Young}, \bibinfo{person}{Nadin
  K{\"{o}}kciyan}, \bibinfo{person}{Isabel Sassoon}, \bibinfo{person}{Sanjay
  Modgil}, {and} \bibinfo{person}{Simon Parsons}.}
  \bibinfo{year}{2018}\natexlab{}.
\newblock \showarticletitle{{Instantiating Metalevel Argumentation
  Frameworks}}. In \bibinfo{booktitle}{{\em COMMA}}. \bibinfo{pages}{97--108}.
\newblock


\bibitem[\protect\citeauthoryear{Zeng, Miao, Leung, and Chin}{Zeng
  et~al\mbox{.}}{2018}]%
        {Zeng2018BuildingArgumentation}
\bibfield{author}{\bibinfo{person}{Zhiwei Zeng}, \bibinfo{person}{Chunyan
  Miao}, \bibinfo{person}{Cyril Leung}, {and} \bibinfo{person}{Jing~Jih Chin}.}
  \bibinfo{year}{2018}\natexlab{}.
\newblock \showarticletitle{{Building More Explainable Artificial Intelligence
  With Argumentation}}.
\newblock \bibinfo{journal}{{\em Thirty-Second AAAI Conference on Artificial
  Intelligence\/}} (\bibinfo{date}{4} \bibinfo{year}{2018}).
\newblock
\showURL{%
\url{https://www.aaai.org/ocs/index.php/AAAI/AAAI18/paper/viewPaper/16762}}


\end{thebibliography}
\end{document}